\documentstyle[epsf]{aipproc}
\begin{document}
\title{Measuring the Higgs CP property 
\\
at a Photon Linear Collider}

\author{Eri Asakawa}
\address{Department of Physics and Graduate School of Humanities
and Sciences, 
\\
Ochanomizu University, Otsuka 2-1-1, Bunkyo-ku,
Tokyo 112-8610, Japan}

\maketitle

\begin{abstract}
We study measurement of the CP property of the Higgs boson
at a photon linear collider.
One method where we take advantage of interference between
Higgs-production and background amplitudes
is proposed. 
A broad peak of the photon energy spectrum 
is helpful in
observing the energy dependence of the interference effects.
Numerical results for the process $\gamma \gamma \rightarrow 
t \overline{t}$ are shown as an example.
\end{abstract}

\section{Introduction}
In the Standard Model (SM), the electroweak
symmetry breaking results in
a physical neutral CP-even Higgs boson.
There also exist various models extended by increasing Higgs fields,
such as the multi-Higgs doublet model where there are extra two neutral
and two charged physical Higgs bosons for each additional doublet.
If CP is a good symmetry in the Higgs sector, 
one of the extra neutral bosons is CP-even and the other
is CP-odd. If CP is not invariant, all of the Higgs states
which have same quantum numbers except for the CP property
can be mixed in their mass eigenstates.
Then, physical Higgs bosons do not have definite CP parity.

One of the notable advantages of a photon linear collider (PLC)
is to provide information 
about the CP property of the Higgs boson by use of linear 
polarizations of colliding photons. 
Defining $J_z = 0$ two-photon states
as $|++ \rangle$ and $|-- \rangle$ where $\pm$ indicate their helicities
with $\hbar$ units,
the CP transformation leads to the other states:
$CP |++ \rangle = |-- \rangle$,
$CP |-- \rangle = |++ \rangle$.
Then, the CP eigenstates are
$(|++ \rangle + |-- \rangle)$ and $(|++ \rangle - |-- \rangle)$;
the former is a $J_z = 0$ component of parallel polarized photons
and couples to a CP-even Higgs boson ($H$)
, and the latter is of perpendicularly polarized photons
and couples to a CP-odd one ($A$).
However,
because colliding beams at PLC are generated by the
Compton back-scattering between high energy electrons and 
laser photons,
the energy spectrum of $\sqrt{s}_{\gamma \gamma}$
distributes broadly and
the degrees of polarizations depend strongly on $\sqrt{\tau}
\equiv \sqrt{s}_{\gamma \gamma}/\sqrt{s}_{ee}$,
where $\sqrt{s}_{\gamma \gamma}$ and $\sqrt{s}_{ee}$
are the center-of-mass energy of $\gamma \gamma$ collisions
and $e^{\pm} e^-$ collisions at parent LC.
In the case of the $500$ GeV LC,
linear polarizations can be used effectively for 
relatively light Higgs bosons whose masses are less than about
a few hundred
GeV. For heavier Higgs bosons, it is necessary that the electron energy
is raised or we use other methods, conventionally.

We propose one method where we take advantage of interference
between Higgs-production and background amplitudes
with circular polarized beams.
A broad peak of the $\sqrt{s}_{\gamma \gamma}$ spectrum 
in the range where the degrees of circular polarizations become large
is helpful to the method, because we have an interest in
the energy dependence of the interference effects.

\section{CP invariant case}
\begin{figure}[b!] 
\begin{center}
\hspace{0cm}
\epsfxsize=7cm
\epsfysize=7cm
\epsffile{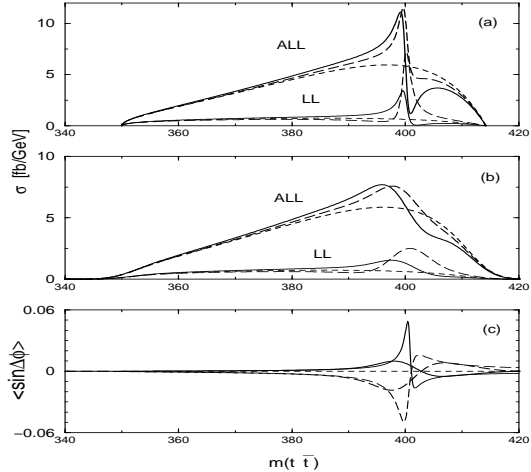}
\vspace{10pt}
\caption{ The $m(t \overline{t})$ dependence of the cross sections
for $\gamma \gamma \rightarrow t \overline{t}$.
The energy and polarization dependence of the $\gamma \gamma$
luminosity has been considered (we set the highest laser frequency,
$x = 4.83$ at $\sqrt{s}_{ee} = 500$ GeV, and $P_e P_L = -1$.).
Short-dashed curves show QED predictions, while solid (long-dashed)
curves show predictions when $A$ ($H$) of $400$ GeV is produced.
In (a) and (b), the thick lines are for total events and the thin lines
are for events where final top-pairs are left-handed.
Gaussian smearing with $\Delta m (t \overline{t}) = 3$ GeV
is applied in (b). Azimuthal decay angular correlation is shown in (c)
with (without) the smearing by thick (thin) lines.} 
\label{fig1}
\end{center}
\end{figure}
As an example,
we consider the process $\gamma \gamma \rightarrow t \overline{t}$
which receives contribution from the Higgs-exchanged $s$-channel
diagram and the top-quark-exchanged $t,u$-channel ones.
The helicity-dependent cross sections are expressed as
\begin{eqnarray}
\frac{d
\sigma^{\lambda_1 \lambda_2 \sigma \overline{\sigma}}}
{d \Omega} \propto
\left| {\cal M}_{\phi}^{\lambda_1 \lambda_2 \sigma \overline{\sigma}}
\right|^2
+
\left| {\cal M}_{cont}^{\lambda_1 \lambda_2 \sigma \overline{\sigma}}
\right|^2
+
2~Re\left[ 
{\cal M}_{\phi}^{\lambda_1 \lambda_2 \sigma \overline{\sigma}}
{\cal M}_{cont}^{\lambda_1 \lambda_2 \sigma \overline{\sigma}}
\right],
\end{eqnarray}
where ${\cal M}_{\phi}$ and ${\cal M}_{cont}$ are the helicity
amplitudes of Higgs resonance ($\phi$ = $H$ or $A$)
and top continuum processes,
$\lambda_{1,2}$ denote the helicities of colliding photons in
$\hbar$ units, $\sigma$ and $\overline{\sigma}$ the final
$t$ and $\overline{t}$ helicities in the center-of-mass frame
in $\hbar/2$ (we also write $-1$ ($+1$) as $L$ ($R$).).

Let us notice interference terms in the above cross sections.
From helicity-dependence of ${\cal M}_{\phi}$~\cite{CPinv},
we find
\begin{eqnarray}
sgn\left[ 
{\cal M}_{\phi}^{\lambda \lambda RR}
{\cal M}_{cont}^{\lambda \lambda RR}
\right]
&=& - sgn \left[ 
{\cal M}_{\phi}^{\lambda \lambda LL}
{\cal M}_{cont}^{\lambda \lambda LL} 
\right] \hspace{5mm} {\rm for} ~~H,
\nonumber \\
sgn \left[ 
{\cal M}_{\phi}^{\lambda \lambda RR}
{\cal M}_{cont}^{\lambda \lambda RR}
\right]
&=& ~~~sgn \left[ 
{\cal M}_{\phi}^{\lambda \lambda LL}
{\cal M}_{cont}^{\lambda \lambda LL} 
\right] \hspace{5mm} {\rm for} ~~A.
\end{eqnarray}
Then, it is possible to distinguish $H$ from $A$ by observing
this difference in the cross section with circular polarized
photons. Numerical results are shown 
in Fig.~\ref{fig1} (a,b).
For definiteness,
we use the MSSM (minimal SUSY SM) prediction for 
the total and partial widths for $A$ adopted in ref.~\cite{CPinv}.
It is found that little interference is observed
for $H$ (long-dashed curves) because the interference effects for
final $t_L \overline{t}_L$ and $t_R \overline{t}_R$ events
cancel each other. On the other hand, the effects for $A$
(solid curves) can be large due to additive interference
for both events.

There is another observable sensitive to CP parity.
When the decay of a top quark
is taken into account,
the cross sections for the processes
$\gamma_{\lambda_1} \gamma_{\lambda_2} \rightarrow 
b W^+ \overline{b} W^-$ can be written by
\begin{eqnarray}
\frac{d
\sigma^{\lambda_1 \lambda_2}_{bW}}
{d \Omega}
\propto 
\sum_{\lambda, \overline{\lambda}} &\biggl\{&
\left| {\cal M}^{\lambda_1 \lambda_2 LL} \right|^2 
\left| {\cal D}_{L}^{\lambda} \right|^2
\Bigl| \overline{\cal D}_{L}^{\overline{\lambda}} \Bigr|^2
+
\left| {\cal M}^{\lambda_1 \lambda_2 RR} \right|^2
\left| {\cal D}_{R}^{\lambda} \right|^2
\Bigl| \overline{\cal D}_{R}^{\overline{\lambda}} \Bigr|^2
\nonumber \\
&+&
2~Re \left[ {\cal M}^{\lambda_1 \lambda_2 LL} 
{\cal M}^{\lambda_1 \lambda_2 RR*} \right]~
Re \left[ {\cal D}_{L}^{\lambda} \overline{\cal D}_{L}^{\overline{\lambda}}
{\cal D}_{R}^{\lambda*} \overline{\cal D}_{R}^{\overline{\lambda}*} \right]
\nonumber \\
&-&
2~Im \left[ {\cal M}^{\lambda_1 \lambda_2 LL} 
{\cal M}^{\lambda_1 \lambda_2 RR*} \right]~
Im \left[ {\cal D}_{L}^{\lambda} \overline{\cal D}_{L}^{\overline{\lambda}}
{\cal D}_{R}^{\lambda*} \overline{\cal D}_{R}^{\overline{\lambda}*} \right]
\biggr\}.
\label{sigma-bw}
\end{eqnarray}
Here, the decay amplitudes for the processes
$t_\sigma \rightarrow b_{-} W^+_{\lambda}$
and
$\overline{t}_{\overline{\sigma}} \rightarrow 
\overline{b}_{+} W^-_{\overline{\lambda}}$
are defined as ${\cal D}_{\sigma}^{\lambda}$ and
$\overline{{\cal D}}_{\overline{\sigma}}
^{\overline{\lambda}}$,
the explicit forms
of which are in the 
appendix of ref.~\cite{D-ron}.
We notice the azimuthal angles of $b$ and $\overline{b}$
in the $t$ and $\overline{t}$ rest frame.
Describing them as $\phi$ and $\overline{\phi}$,
they appear in the third and fourth terms
in Eq.~\ref{sigma-bw}:
\begin{eqnarray}
Re \left[ {\cal D}_{L}^{\lambda} \overline{\cal D}_{L}^{\overline{\lambda}}
{\cal D}_{R}^{\lambda*} \overline{\cal D}_{R}^{\overline{\lambda}*} \right]
\propto \cos(\phi- \overline{\phi}),
\nonumber \\
Im \left[ {\cal D}_{L}^{\lambda} \overline{\cal D}_{L}^{\overline{\lambda}}
{\cal D}_{R}^{\lambda*} \overline{\cal D}_{R}^{\overline{\lambda}*} \right]
\propto \sin(\phi- \overline{\phi}).
\end{eqnarray}
Therefore, we obtain
\begin{eqnarray}
\langle \sin(\phi-\overline{\phi}) \rangle = \frac
{{\displaystyle\int} 
\sin(\phi-\overline{\phi}) ~{\displaystyle\frac{d\sigma
^{\lambda_1 \lambda_2}_{bW}
}{d\Omega}} ~d\Omega}
{\displaystyle \int {\frac{d\sigma
^{\lambda_1 \lambda_2}_{bW}
}{d\Omega}} ~d\Omega}
\propto
Im \left[ {\cal M}^{\lambda_1 \lambda_2 LL} 
{\cal M}^{\lambda_1 \lambda_2 RR*} \right].
\end{eqnarray}
When colliding photons are polarized to be $+1$,
$Im \left[ {\cal M}^{++ LL} {\cal M}^{++ RR*} \right] \simeq
Im \left[ {\cal M}^{++ LL}_{\phi} {\cal M}^{++ RR*}_{cont} \right]$
is satisfied considering $\left| {\cal M}_{cont}^{++RR} \right|
\gg \left| {\cal M}_{cont}^{++LL} \right|$.
Since the Higgs-production amplitudes ${\cal M}^{++ LL}_{H}$ and 
${\cal M}^{++ LL}_{A}$ have opposite signs in the MSSM-type models,
$\langle \sin(\phi-\overline{\phi}) \rangle$ tells us the
CP parity of the Higgs boson.
A numerical calculation is shown in Fig.~\ref{fig1}(c). Though
the quantity turns out to be rather small because of cancellation
between the contributions from longitudinally and transversely
polarized $W$'s, we can recover the sensitivity to 
the CP parity by taking account of $W$-decay distributions.

\section{CP non-invariant case}
When the CP symmetry is not conserved in the Higgs potential,
the helicity amplitudes for 
the Higgs-production are denoted by
\begin{eqnarray}
{\cal M}_\phi^{\lambda_1\lambda_2 \sigma\overline{\sigma}}
   = \frac{e\,\alpha}{4\pi} \frac{m_t}{m_W}
     \frac{s}{s-m_\phi^2+im_\phi\Gamma_\phi} 
     \left[\,a_\gamma + \lambda_1 b_\gamma \right]
     \left[\,\sigma\beta_t a_t -i b_t\right]
     \delta_{\lambda_1 \lambda_2}\delta_{\sigma \overline\sigma}\,.
\label{mephi}
\end{eqnarray}
where $a_{\gamma}$ and $a_t$ are proportional to the CP-even components
of vertices, $\phi \gamma \gamma$ and $\phi t \overline{t}$, 
$b_{\gamma}$ and $b_t$ to the CP-odd components.
Since \{$a_{\gamma}, b_{\gamma}$\} or/and \{$a_t, b_t$\} 
have non-zero values
simultaneously, they induce complicated interference.
Moreover, the amplitudes include six parameters for vertices;
$a_{\gamma}, b_{\gamma}$ are complex, whereas
$a_t, b_t$ are real.
Therefore, we need at least six observables to determine the CP property
completely and are urged to use linear polarizations as well.
All observables obtained from various polarizations (including
the mixture of linear and circular ones) are exhibited 
in ref.~\cite{CPninv}.

\section{Conclusions}
We have discussed measurement of the CP property of the Higgs boson
at PLC. It has been found that
we can extract information about the CP property of
the Higgs boson from the observation of interference effects
between Higgs-production and background amplitudes.
If the Higgs boson have definite CP parity,
this method can be powerful in high $\sqrt{\tau}$ region
where linear polarizations of colliding photons become useless
conventionally.

\vspace{3mm}
I would like to thank K.~Hagiwara for valuable discussions
and G.C.~Cho for helpful comments.

\end{document}